# Genetic Code Table: A note on the three splittings into amino acid classes

Miloje M. Rakočević

*Faculty of Science, University of Niš (now retired, on the Address: Milutina Milankovica 118, 11070 Belgrade, Serbia (e-mail: m.m.r@eunet.yu; or: milemirkov@nadlanu.com; www.sponce.net or www.rakocevcode.rs)*

**Abstract**

This note represents the further progress in understanding the determination of the genetic code by Golden mean (Rakocevic, 1998). Three classes of amino acids that follow from this determination (the 7 "golden" amino acids, 7 of their complements, and 6 non-complements) are observed now together with two further possible splittings into 4 x 5 and 5 x 4 amino acids.

In one of the previous works (Rakočević, 1998) was shown that the Genetic Code Table (GCT) of the standard genetic code (Table 1) can be developed in a six-bit binary-code tree (Figure 1). Thus, the order of the eight codon octets is: YUN, YCN, RUN, RCN, YAN, YGN, RAN, RGN. [One-letter abbreviations: Y from pYrimidine; R from puRine; and N from aNy (of bases).] (see Comment 1).

With respect to the Golden mean (Table 2), in such a binary tree one can establishe the positions of amino acids (AAs). As result of the Golden mean determination there are three classes of AAs: the 7 "golden" AAs, the 7 their complements (complements to the full pairs) and the 6 non-complements, all as in Table 3. Then the number of atoms (60, 66 and 78) within the three classes differs for 1 x 6; 2 x 6 and 3 x 6 of atoms, respectively (see Comment 2).

In this note, however, we show that the determination with the number 6 (the first perfect number!)[1] has new aspects related to positions (odd/even) of AAs within the AAs classes (Tables 4-9).

[1] The factors of perfect numbers are in correspondence to the binary sequence $2^n$ (n = 0, 1, 2, 3, ...) as follows. For the first (6): 1, 2, (4 -1 = 3), (8 – 2 = 6); for the second (28): 1, 2, 4, (8 -1 = 7), (16 – 2 = 14), (32 – 4 = 28); for the third (496): 1, 2, 4, 8, 16, (32 -1 = 31), 64 – 2 = 62), (128 – 4 = 124), (256 – 8 = 248) , (512 – 16 = 496); for the forth (8128): 1, 2, 4, 8, 16, 32, 64, (128 -1 = 127), (256 –

If the first two classes of AAs (in the upper part of Table 3) are seen as one class, then it can be reclassified into two new classes, depending on the question – whether AAs are on the even, or on the odd positions; all these in the order dictated by GCT, or by the hierarchy of number of atoms in the amino acid molecules (their side chains), either within the set of "golden" amino acids or their complements.

The separation of the third class of AAs (in the lower part of Table 3) has its own chemical as well as arithmetical justification. Chemically, the four aliphatic AAs (D, E & K, R)[2] are specific in relation to all other aliphatic AAs, because two of them contain the same functional group (carboxylic) twice – in the "head" and in the "body" of an amino acid molecule. The same is valid for K and R, but in relation to the basic amine functional group, $NH_2$. (Notice that within N and Q there is an amide functional group.)

For two aromatic AAs (H and W) it can not be said that they contain the same functional group just twice, but it can be said that they contain a double function (aromatic and heterocyclic), while the remaining two aromatic AAs (F and Y) contain only a single function (aromatic).

This sample of chemical hierarchy has a match in arithmetical hierarchy. So, while the two first classe determination by number 6 appears only once (10 x 6 and 11 x 6, respectively), so far for the third class the determination is seen by twice: the original determination with 13 x 6 is "split" into two determinations: 5 x 6 and 8 x 6 (see Comment 3). Really, within four aliphatic AAs (D, E & K, R) there are (8 x 6)+1, whereas within two aromatic, H & W, one has (5 x 6) -1 of atoms.

The Table 10 contains a splitting into three amino acid classes (according to Table 3), while Table 11 contains a new splitting into five classes: one class of aromatic and four classes of aliphatic AAs. If, however, in Table 11 we join aromatic AAs to aliphatic ones, one gets four classes: the first (F) to the first class, the second (Y) to the second one (going from top to down); and also the first (H) to the first class, the second (W) to the second one (going from bottom to up). The number of atoms within obtained four classes is also determined by the multiples of number 6, with the deviations for ±0 and/or ±1 (Table 12). Particularly, it is interesting the possibility of a new splitting also into the three classes as it is shown in Table 13. In relation to Table 3 we have a change for ± 1 and/or ± 0 molecules as well as for the ± 1 and/or ± 0 of atoms within the amino acid classes.

---

2 = 254), (512 – 4 = 508), (1024 – 8 = 1016), (2048 -16 = 2032), 4096 – 32 = 4064), (8192 – 64 = 8128): etc. Here one must notice that there is a connection with the friendly numbers through third friendly number, i.e. through the first member of second friendly pair (1184), and a connection with Shcherbak's system of multiples of number 037, valid for GC (Shcherbak, 1994, Figure 1 and Table 1) at the same time: 1, 2, 4, 8, 16, 32 / 37, 74, 148, 296, **592**, 1184 (see position of the number **592**, as the 16th case in Table A.1, in Appendix). The first three friendly pairs are: (220, 284), (1184, 1210), (17296, 18416).

[2] Starting from this chemically strict defined class of AAs it makes sense all the other "quadruplets" also seen as possible chemical classes, what means a splitting into five amino acid classes (5 x 4) as it will be showed in further text.

From the original "Golden Table" (Table 3), taking the sequence of "golden" AAs in molecule mass ordering, i.e. in hierarchy over atom number (Table 9), follows still one new amino acid splitting – the splitting into 5 x 4 AAs (Table 14).

In the middle position of this Cyclic Invariant Periodic System (CIPS)[3], presented in Table 14, there are chalcogene AAs (S, T & C, M)[4], and then in a cyclic arrangement follow the "contact" AAs (G, P & V, I)[5], two double acidic AAs with two their amide derivatives (D, E & N, Q)[6]; the two original aliphatic AAs with two amine derivatives (A, L & K, R)[7]; and, finally, four aromatic AAs (F,Y & H, W). Notice that within CIPS each amino acid position is strictly determined and none can be changed.

As it is immediately obvious from CIPS, the nature of the genetic code again points out the validity of Aristotle sentence[8], and, on the other hand, necessarily

---

[3] Cyclicity and periodicity through the positions of two and two amino acids – up/down – in relation to middle chalcogene AAs.

[4] The elements from the sixth group of Periodic system are the chalcogenes, and that is valid for oxygen and sulfur. Hence follows that these AAs are just such – the chalcogene AAs! Of course, for still two amino acids (D and E), it can be concluded that from their nature, they are also chalcogene AAs. However, when we see that here is in the question a system *per se*, a system of "5 x 4" elements, then we can say that four amino acids (S, T & C, M) are the source chalcogene AAs, and the other two (D, E) are derived; they must go into a separate class with its derivatives, two amides: N, and Q. [Cf. legend to Table B. 9, where it is shown that (S, T & C, M) and (D, E) are together.]

[5] After Popov (Popov 1989; Rakočević & Jokić, 1996) the four "contact" AAs (G, P & V, I) are of the non-alanine stereochemical types: G from glycine type, P from proline, and V, I from valine stereochemical type. The term "contact" is our , and the explanation is in Comment 4 (Rakočević, 2006).

[6] The amino acids D & E are double acidic because carboxylic gropu exists in both „head" and „body" of molecule at the same time.

[7] The source, i.e. original aliphatic compounds are nonpolar; their amine derivatives are less polar then hydroxy derivatives ($NH_2$ less polar than OH, and N less than O).

[8] "The existence of such a harmonic structure with unity of a determination with physical–chemical characteristics and atom and nucleon number ... appealed to Aristotle and to his idea of unity of form and essence" (Rakočević, 2004, p. 233). From the aspect of this paper, the form make the

leads to the conclusion that all three main hypotheses about the origin of the genetic code are *mutatis mutandis* valid; two given by Crick (the genetic code as a result of stereochemical interactions[9], or as a result of pure chance[10]) (Crick, 1968) and one by Wong, the theory of co-evolution[11]. (Wong, 1975, p. 1909: "The theory is proposed that the structure of the genetic code was determined by the sequence of evolutionary emergence of new amino acids within the primordial biochemical system".)[12]

As we can see from Tables which follow from CIPS (see Appendix B), the key principles valid for this system are: 1. Principle of minimum change, 2. Principle of simple proportion, 3. Principle of multi meaning diversity, 4. Block-aufbau principle, and 5. Principle of self-similarity. All together these are the principles of building of specific harmonic structures[13] or, shorter, the harmonic principles.

---

Golden mean determined amino acid positions within CIPS, and the essence is realized through two realities: physically, through molecule mass; and chemically – through the expression of the five revealed chemical classes (see Comment 5).

[9] Crick, 1968, p. 369: "This theory states that the code is universal because it is necessarily the way it is for stereochemical reasons".

[10] Crick, 1968, p. 370: "In its extreme form, the theory implies that the allocation of codons to amino acids at this point was entirely a matter of 'chance' ". For the aspect of CIPS existence that means that distribution of codons to amino acids must be in relation to a pure form, such as it is Golden mean; then, in relation to size of amino acid molecule and in relation to physical–chemical properties at the same time; by this, it is matter of a pure chance to "find" such "golden" conditions in a prebiotic area (see Comment 6).

[11] The act of a "co-evolution" itself is not possible without a "co-influence", i.e. without an interactive influence of amino acid and/or nucleotide components of genetic code (Rakočević, 2004, p. 233: "The word can be only about a 'co-evolution' based on 'co-influence' of more factors").

[12] On a slightly different way, the same idea is expressed by Sukhodolec (Sukhodolec, 1981, p. 499: „The basis of the hypothesis is referred to the idea that during the prebiotic evolution amino acids and nitrogenous bases existed in the form of complex crystal structures and that in the construction of these structures, as components of their elements, the chain units - analogues alternately have been used: the amino acids and duplexes from the first two bases of the codons". Our idea is similar to this idea of Sukhodolec (Rakočević, 2004, p. 232: "At a later stage many nucleotide/amino-acid aggregations, similar to aggregations of Miller's type, or to not much different aggregations of Murchison–meteorite's type, had been realized").

[13] The structures like this one, presented in our previous paper (Rakočević, 2004) (see Table D.3).

| 1st lett. | 2nd letter: U | 2nd letter: C | 2nd letter: A | 2nd letter: G | 3rd lett. |
|---|---|---|---|---|---|
| U | 00. UUU<br>01. UUC **F**<br>02. UUA<br>03. UUG **L** | 08. UCU<br>09. UCC<br>10. UCA **S**<br>11. UCG | 32. UAU<br>33. UAC **Y**<br>34. UAA<br>35. UAG CT | 40. UGU<br>41. UGC **C**<br>42. UGA CT<br>43. UGG **W** | *U*<br>*C*<br>*A*<br>*G* |
| C | 04. CUU<br>05. CUC<br>06. CUA **L**<br>07. CUG | 12. CCU<br>13. CCC<br>14. CCA **P**<br>15. CCG | 36. CAU<br>37. CAC **H**<br>38. CAA<br>39. CAG **Q** | 44. CGU<br>45. CGC<br>46. CGA **R**<br>47. CGG | *U*<br>*C*<br>*A*<br>*G* |
| A | 16. AUU<br>17. AUC **I**<br>18. AUA<br>19. AUG **M** | 24. ACU<br>25. ACC<br>26. ACA **T**<br>27. ACG | 48. AAU<br>49. AAC **N**<br>50. AAA<br>51. AAG **K** | 56. AGU<br>57. AGC **S**<br>58. AGA<br>59. AGG **R** | *U*<br>*C*<br>*A*<br>*G* |
| G | 20. GUU<br>21. GUC **V**<br>22. GUA<br>23. GUG | 28. GCU<br>29. GCC **A**<br>30. GCA<br>31. GCG | 52. GAU **D**<br>53. GAC<br>54. GAA **E**<br>55. GAG | 60. GGU<br>61. GGC **G**<br>62. GGA<br>63. GGG | *U*<br>*C*<br>*A*<br>*G* |

**Table 1.** The Table of the standard genetic code. Ordinal number of codons after the order-key: YYN, RYN, YRN, RRN, in correspondence with the hierarchy on the binary-code tree in Figure 1.

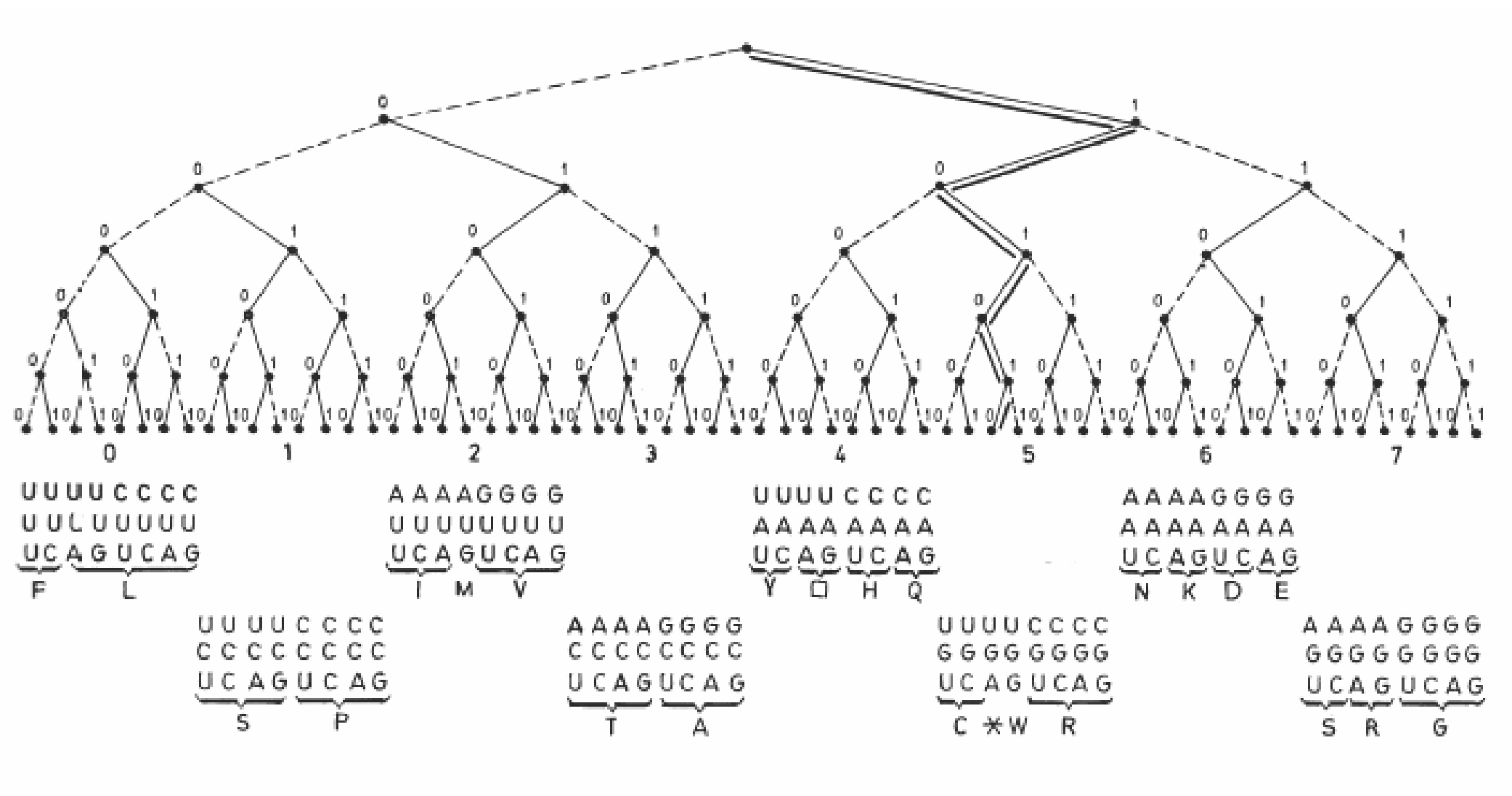

**Figure. 1**. Genetic code as a binary-code tree. The full lines: the routes of the greater changes, from 0 to 1 and vice versa; the dotted lines: the routes of the less changes, from 0 to 0, as well as from 1 to 1 going from a higher into a lower level. The double full line: the route of the maximum possible changes: from 0 to 1 and vice versa in any step (the route corresponding to the ‘Golden mean route’ on the Farey tree, Figure 2). Asterisks: ‘stop’ codon UGA; square: ‘stop’ codons UAA and UAG. The codon distribution and order after the rules given by R. Swanson (Swanson, 1984, Figure 1): 1 for purine and 0 for pyrimidine; 1 for three and 0 for two hydrogen bonds.

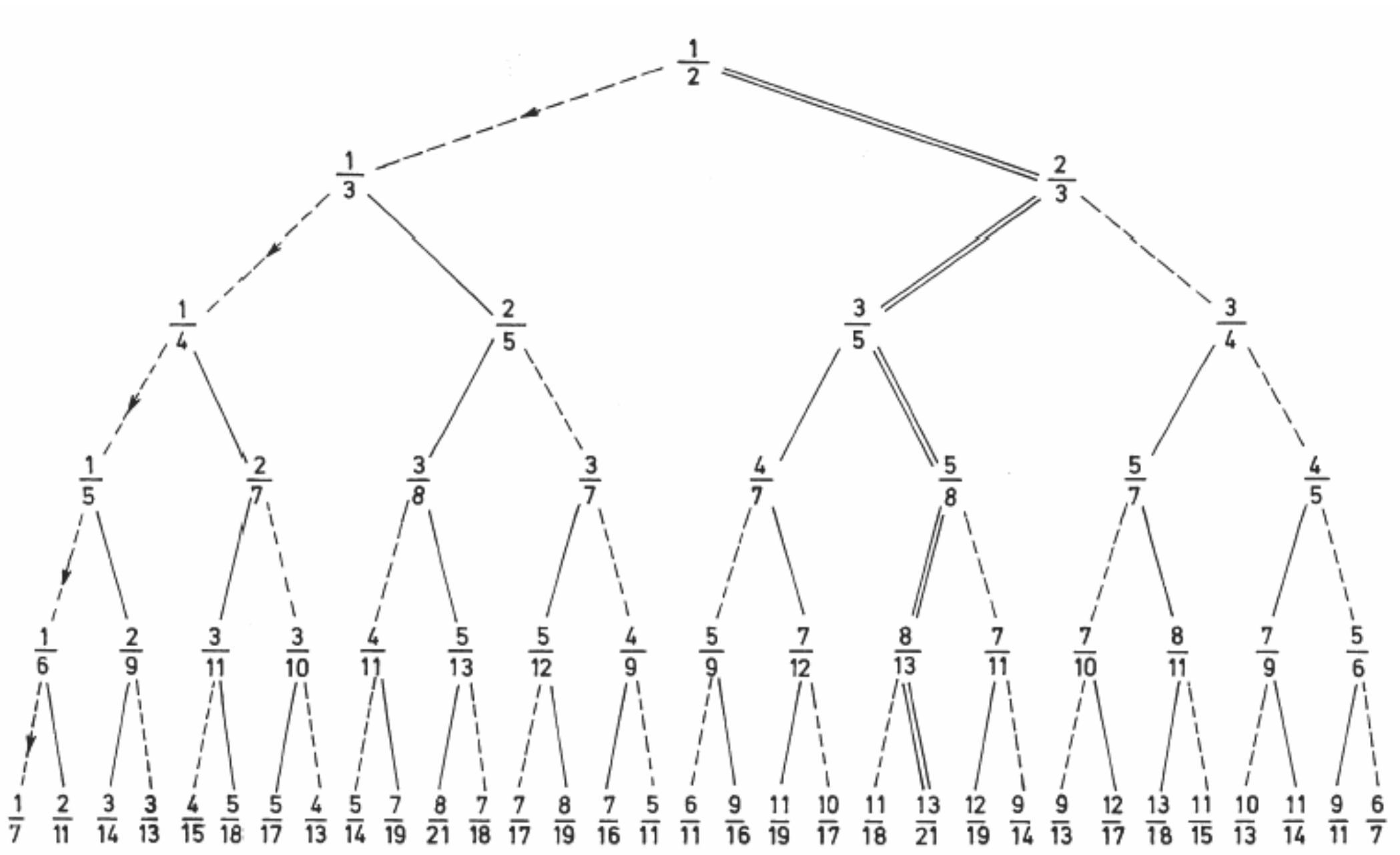

**Figure 2**. The Farey binary tree. It has a special application in physics of the deterministic chaos, and theory of fractals. At the same time at this tree it is possible to give a proof that the set of rational numbers is countable. The double line represents the "golden route" because it possesses Fibonacci numbers (related to Golden mean) in numerators as well as denominators. [This graph is the same as in original paper (Rakočević, 1998) with a minor error: the line 3/5 – 4/7 must be a dotted one.]

| $\phi^0$ | $\phi^1$ | $\phi^2$ | $\phi^3$ | $\phi^4$ | $\phi^{5-7}$ | $\phi^8$ | $\phi^9$ |
|---|---|---|---|---|---|---|---|
| G | Q | T | P | S | L | L | F |
| 63 | 39-38 | 25-24 | 15-14 | 10-09 | 06-02 | 02-01 | 01-00 |
| 63 | 38.94 | 24.06 | 14.87 | 9.19 | 5.68 – 2.17 | 1.34 | 0.83 |

**Table 2.** The amino acids in Golden mean power positions within the sequence 0–63 on the binary-code tree in Fig. 1. First row: Golden mean powers within first 'cycle' in module 9. Second row: amino acids in the positions marked in third row, taken from the binary-code tree in Fig. 1. Fourth row: the values of the Golden mean powers within the interval 0–63. The calculations: 0.618033 x 63 = 38.94; 0.618033 x 0.618033 x 63 = 24.06 etc.

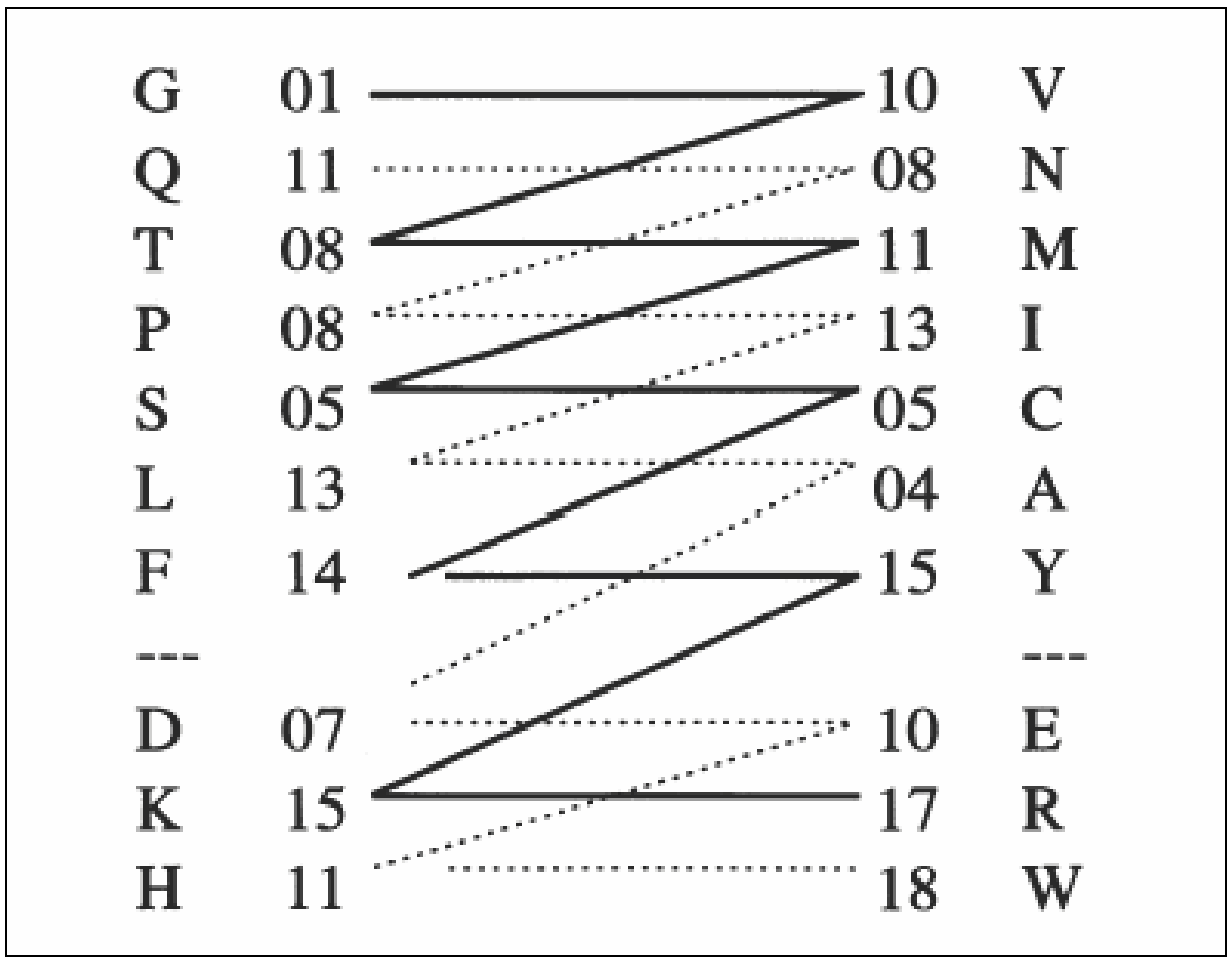

**Table 3.** Atom number balance directed by Golden mean on the binary-code tree (Scheme 2 in Rakočević, 1998). First seven amino acids on the left are ‘golden’ amino acids with 60 atoms within side chains; on the right are their (chemically) pairing complements with [60 + (1 x 6)] = 66 atoms; below are three amino acid pairs as non-complements with [66 + (2 x 6)] = 2 x 39 = 78 atoms; Notice that within aliphatic non-complements there are 39+10, whereas within aromatics (H & W) 39-10 of atoms; in the other words: (8 x 6) +1 and (5 x 6) -1, respectively. On the first zigzag (full) line there is 102-1 whereas on the second (dotted) line 102+1 atoms. Arithmetic mean for both: 102±1. Notice also that the arrangement-ordering of “golden” AAs is the same as in Table 2.

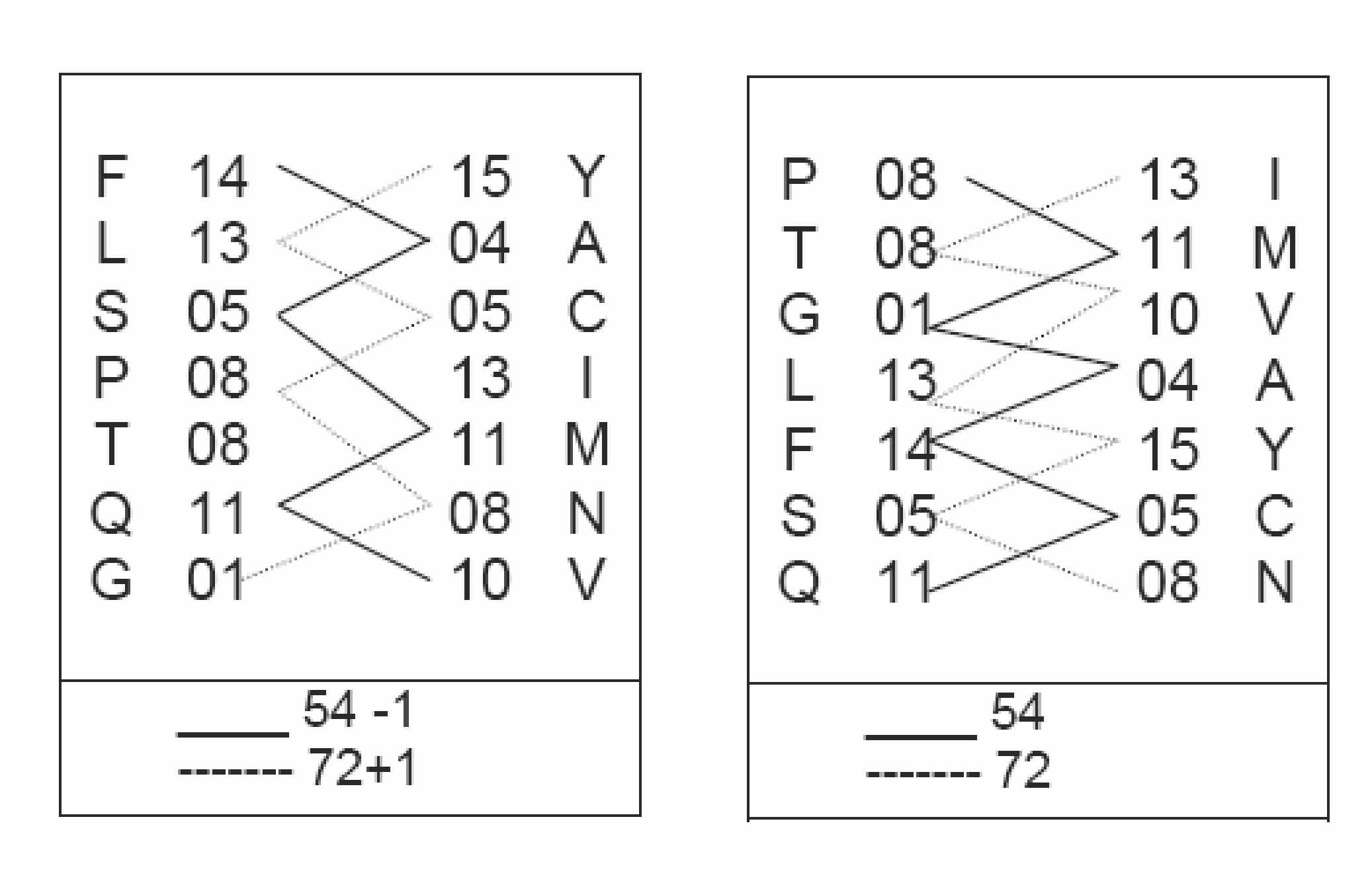

**Table 4 (left).** The atom number balance within two classes of AAs. Both classes as in Table 3, in a reverse order, from the aspect of the position of "up-down"; there: G, Q, T, P, S, L, F; and here: F, L, S, P, T, Q, G. On the first zigzag (full) line there are [(9 x 6)-1] whereas on the second one (dotted line) [(12 x 6) +1] of atoms.

**Table 5 (right).** The atom number balance within two classes of AAs: all the same as in Table 4 except the order of AAs is given according to the positions of complements (AAs on the right) on the binary tree, as well as in GCT; on the first zigzag (full) line there are exactly 9 x 6 whereas on the second one (dotted line) 12 x 6 of atoms. (Cf. Table A.5, where aromatic AAs are included with a new type of balance.)

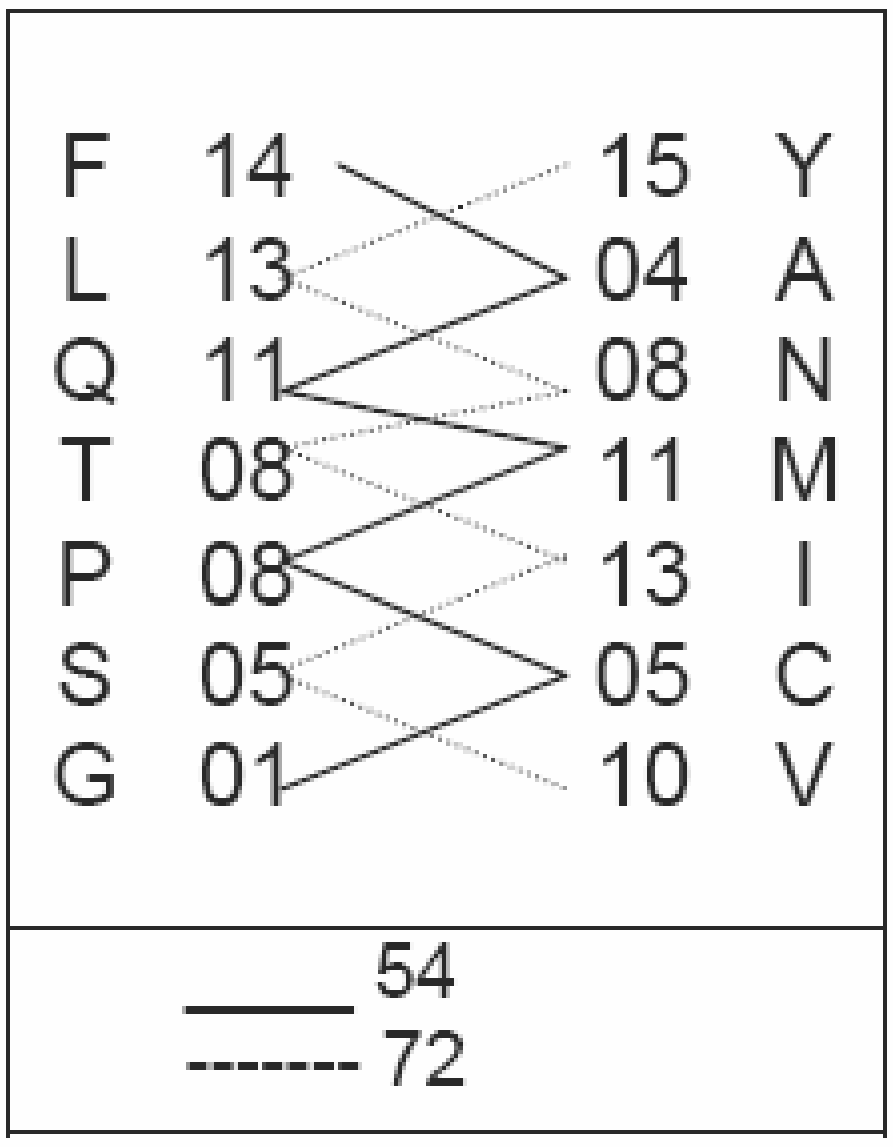

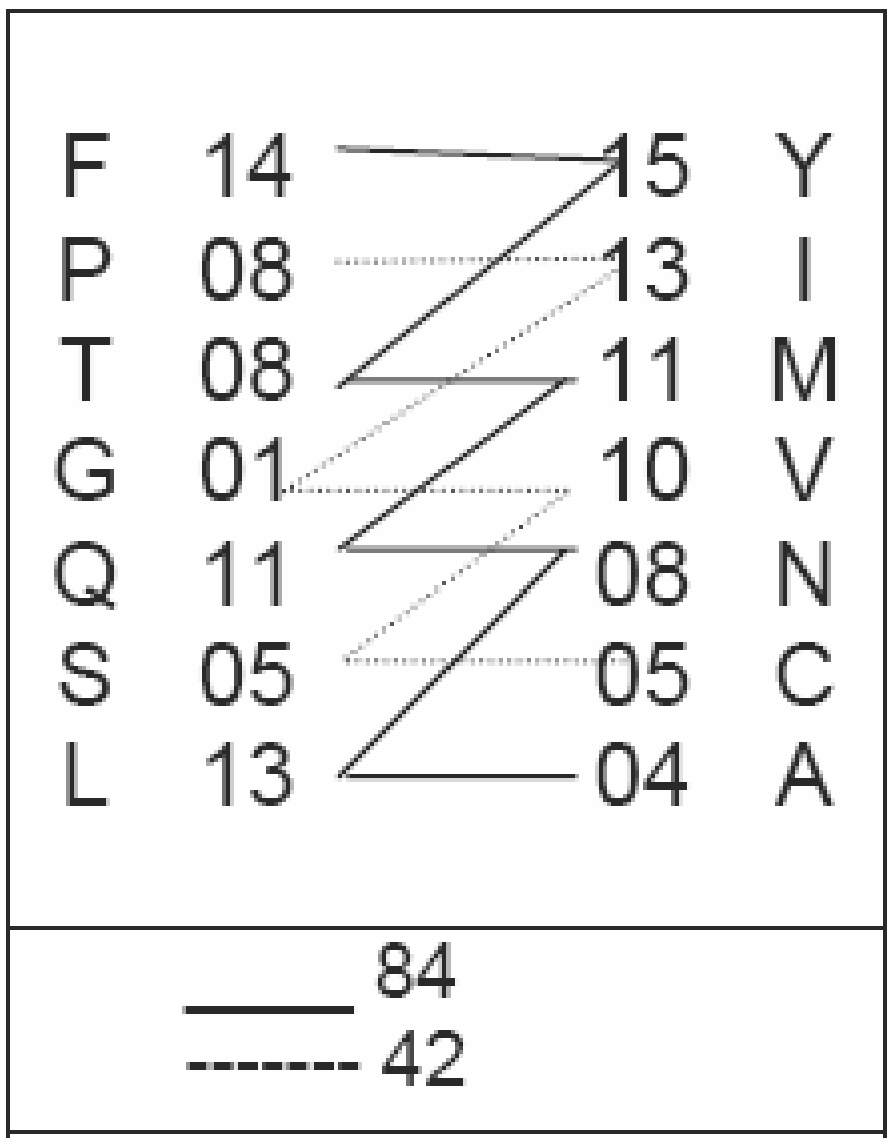

**Table 6 (left).** The atom number balance within two classes of AAs. Both classes as in Tables 4 and 5, but the order of AAs is given according to the number of atoms in side chains of "golden" AAs (AAs on the left). On the first zigzag (full) line there are exactly 9 x 6 whereas on the second one (dotted) 12 x 6 of atoms.

**Table 7 (right).** The atom number balance within two classes of AAs. Both classes as in Tables 4 and 5, but the order of AAs is given according to the number of atoms in side chains of "complement" AAs (AAs on the right). On the first zigzag (full) line there are exactly 14 x 6 whereas on the second one (dotted) 7 x 6 of atoms; both lines in proportion 2:1 what is "the symmetry in the simplest case" (Marcus, 1989).

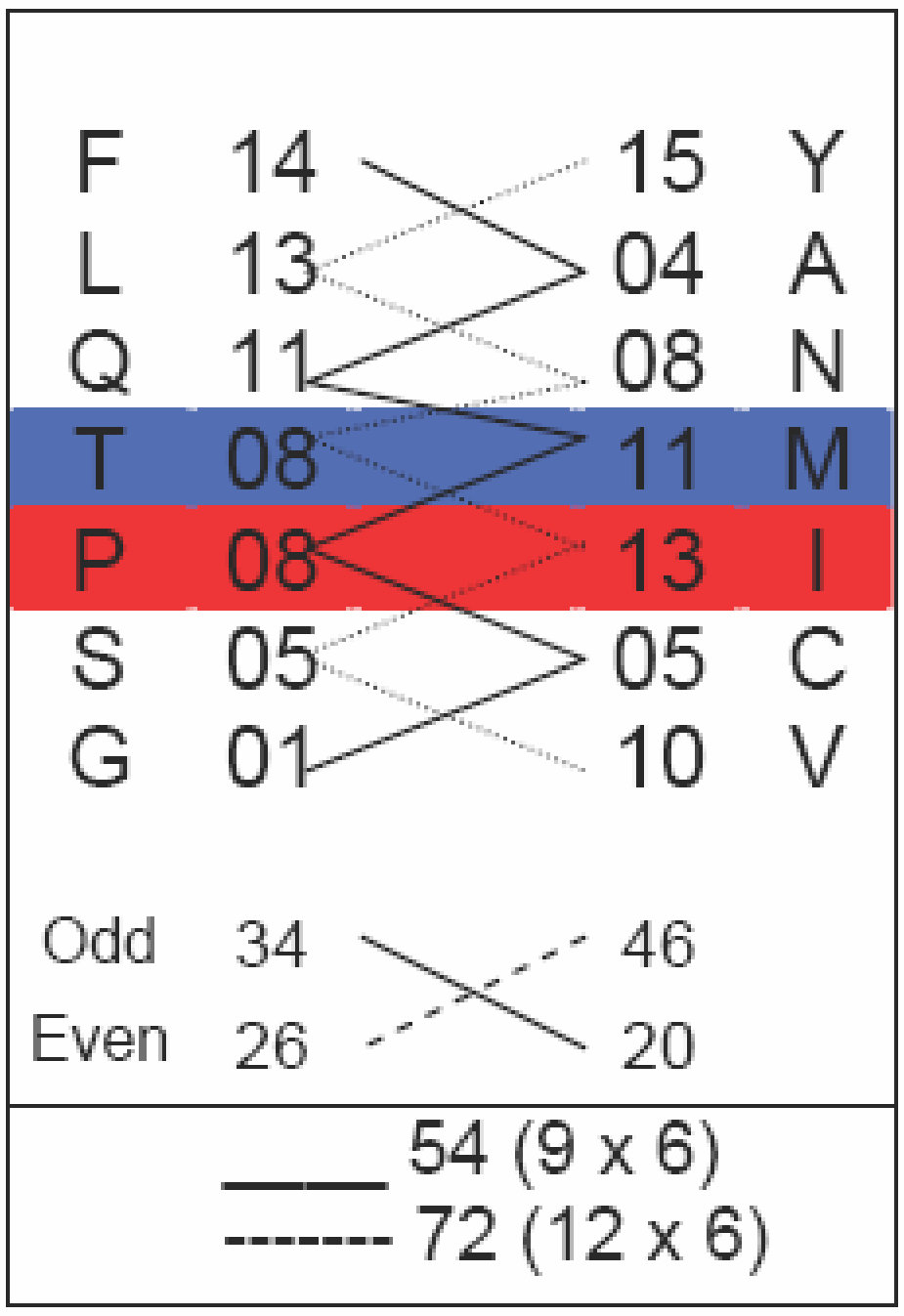

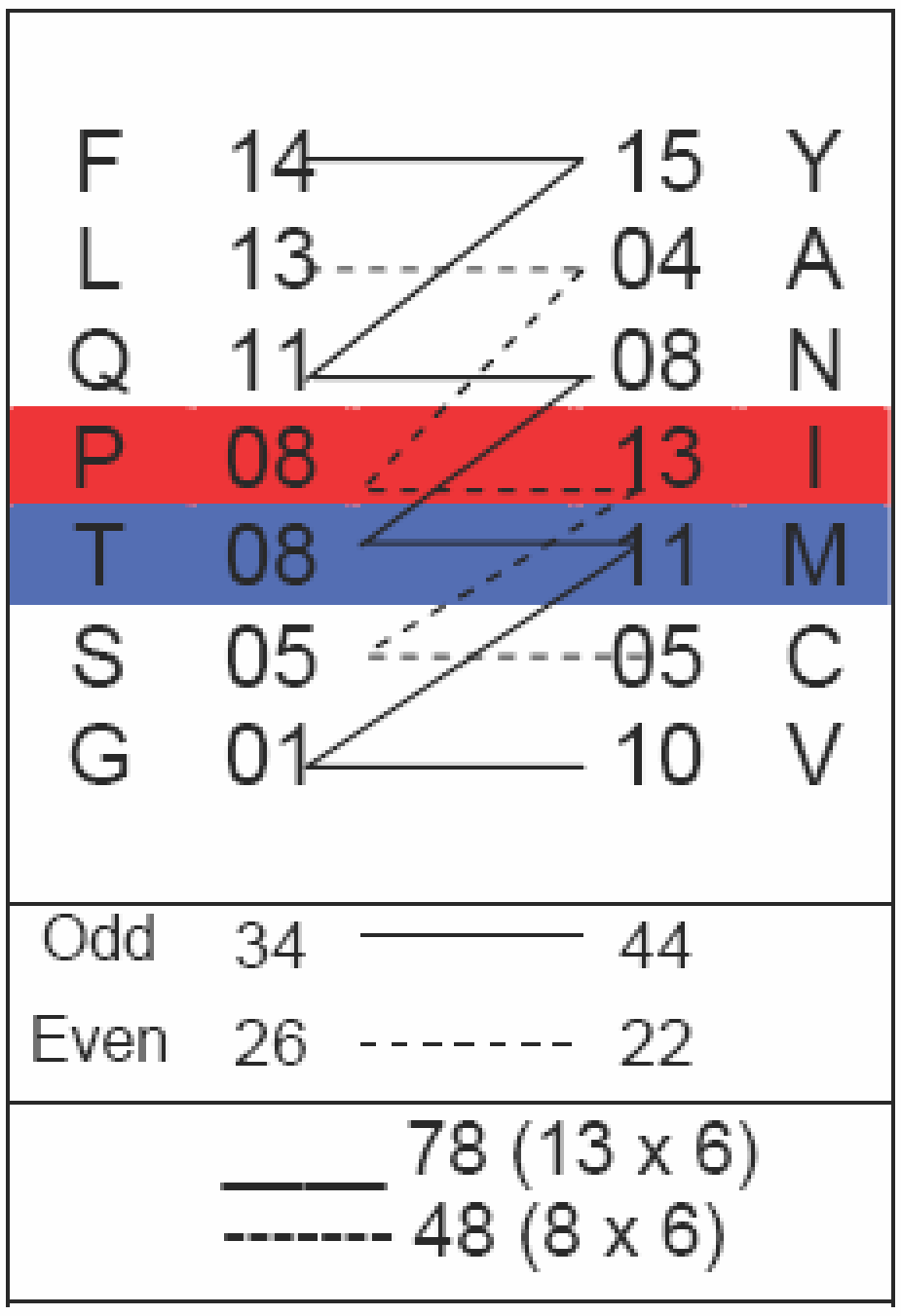

**Table 8 (left).** The Table as Table 6, but here with the results of calculation the number of atoms at even and odd positions, and with the mark of positions of P and T.

**Table 9 (right).** The Table as Table 8, but with a change of the position of P in relation to the T; so the S and the T are now in the contact, i.e. in neighborhood. For P is taken that it possesses 8 and not 9 atoms within the side chain because the "head" must be taken with a complete amine ($NH_2$) group, as by Shcherbak (1994) in nucleon number calculation. Realistically perceived, the order of P-08 here is determined by the order of his mate I-13.

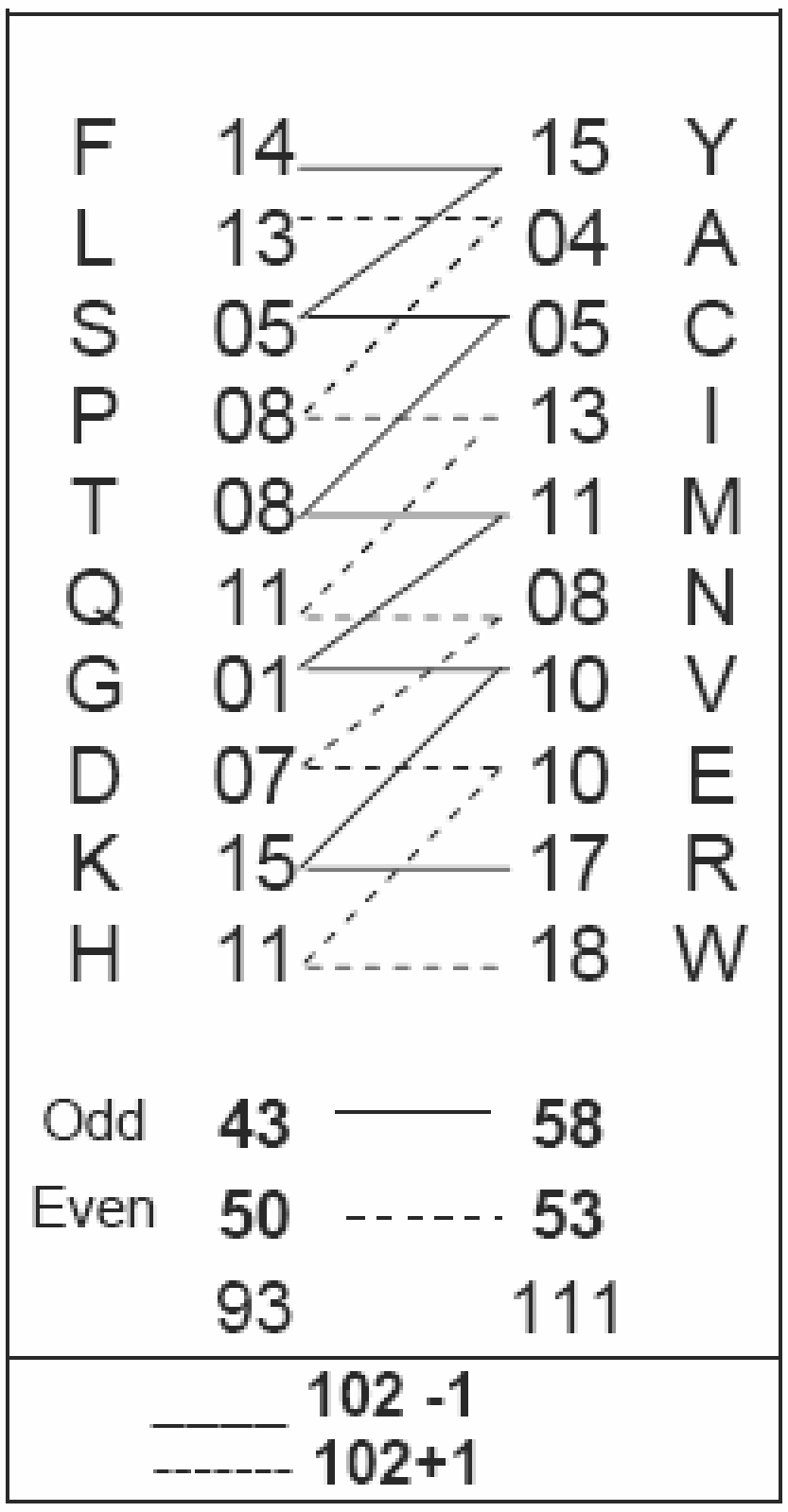

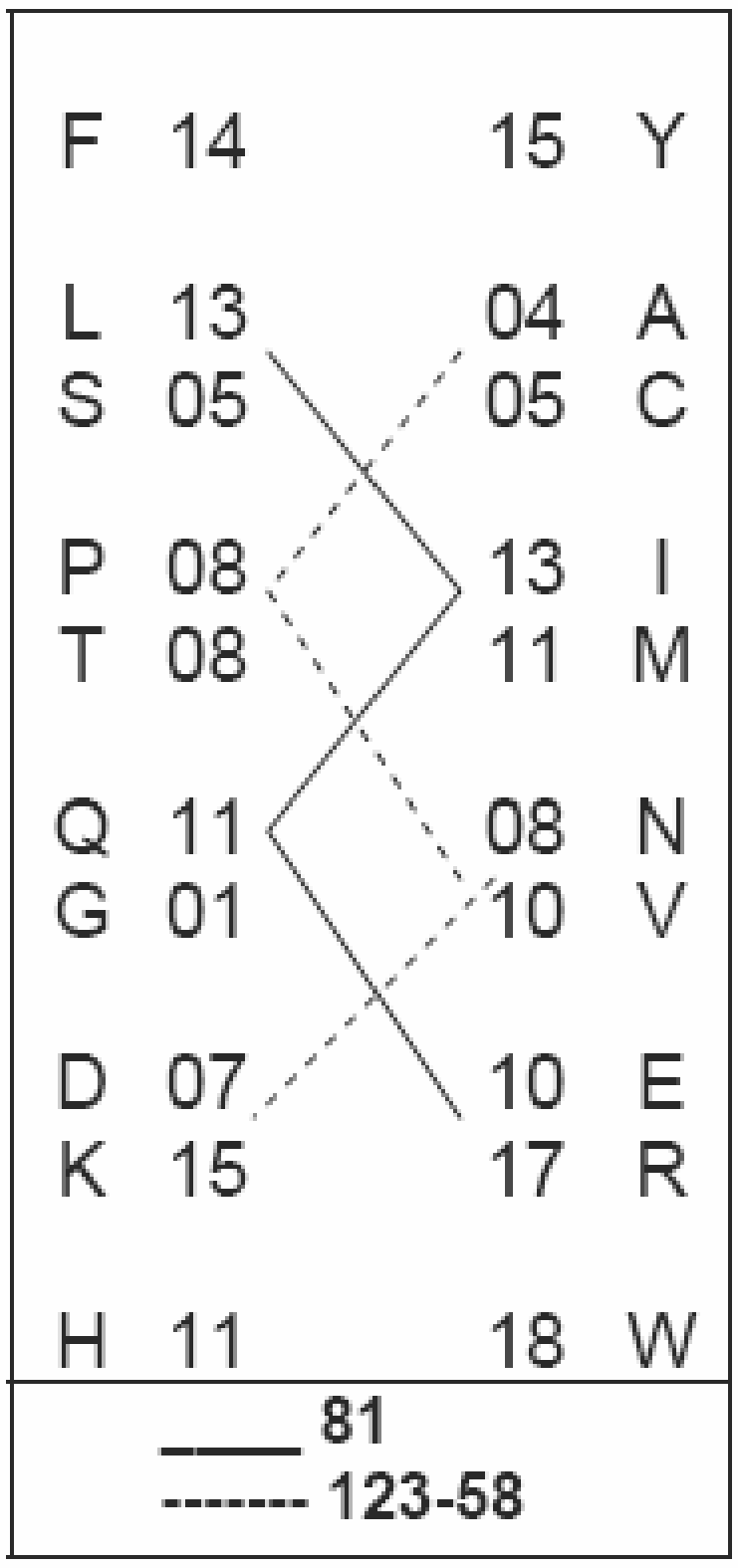

**Table 10 (left).** This Table is the same as Table 3 in all, except the sequence of "golden" AAs (as well as their complements) is in a vice versa order; still, the results of atom number calculations at even/odd positions. Notice that within aliphatic AAs on the left (column) there is 68 x 1, whereas on the right 78 atoms. Notice also that 78 + 58 (in aromatics) equals 68 x 2 (cf. Table B.2).

**Table 11 (right).** The same Table as Table 10, but given in amino acid pairs within the "quadruplets". The balance for aliphatic AAs corresponds to the balance valid for two amino acid classes handled by two classes of synthetases (Tables D.1 and D.2) [81 versus 123 atoms, where the aromatic AAs (their side chains) possess 58 of atoms]

| | | | | |
|---|---|---|---|---|
| F | 14 | | | |
| L | 13 | (42-1) | 04 | A |
| S | 05 | (7x6) | 05 | C |
| | | | 15 | Y |
| P | 08 | (54+1) | 13 | I |
| T | 08 | (9x6) | 11 | M |
| Q | 11 | (48±0) | 08 | N |
| G | 01 | (8x6) | 10 | V |
| | | | 18 | W |
| D | 07 | (60±0) | 10 | E |
| K | 15 | (10x6) | 17 | R |
| H | 11 | | | |
| (17 x 06 = 102) ±1 (7+10 = 8+9 = 17) | | | | |
| F | 14 | | 15 | Y |
| L Q | 24 | (65) | 12 | A N |
| P T | 16 | | 24 | I M |
| | | (61) | | |
| S G | 06 | | 15 | C V |
| D K | 22 | (78) | 27 | E R |
| H | 11 | | 18 | W |
| (66 -1) / (60+1) / 78±0 | | | | |

**Table 12 (left).** Table follows from Table 11 joining aromatic AAs to aliphatic ones: F and Y up, H and W down. Within two outer quintets (of dark tones) there are (17 x 06) - 1, whereas within two inner with light tones (17 x 6) + 1 of atoms.

**Table 13 (right).** Table is generated from Table 9 with an adding of non-complements and a new amino acid grouping. In relation to Table 3, instead 60, 66, 78 atoms here are (60+1), (66-1) and 78±0 of atoms.

| | | | | | |
|---|---|---|---|---|---|
| 073 | F | 14 | 15 | Y | 079 |
| 235 | L | 13 | 04 | A | 172 |
| 087 | Q | 11 | 08 | N | 085 |
| 160 | P | 08 | 13 | I | 121 |
| 168 | T | 08 | 11 | M | 043 |
| 243 | S | 05 | 05 | C | 081 |
| 184 | G | 01 | 10 | V | 168 |
| 087 | D | 07 | 10 | E | 093 |
| 091 | K | 15 | 17 | R | 265 |
| 081 | H | 11 | 18 | W | 044 |

**Table 14.** The Cyclic Invariant Periodic System (CIPS) of canonical AAs. At the outer side, left and right, it is designated the number of atoms within coding codons; more exactly, in the Py-Pu bases (U = 12, C = 13, A = 15 and G = 16) (cf. Table 3 in Rakočević, 1997b, p. 648); at the inner side – the atom number within amino acid side chains (see Tables B.4 – B.9). In the middle position there are chalcogene AAs (S, T & C, M); follow - in next „cycle“ - the “contact” AAs (G, P & V, I), then two double acidic AAs with two their amide derivatives (D, E & N, Q), the two original aliphatic AAs with two amine derivatives (A, L & K, R); and, finely, four aromatic AAs (F,Y & H, W) – two up and two down.

## COMMENTARIES

1. The order of codons in Table 1 corresponds to the chemical complexity of molecules: purine, as imidazole derivative of pyrimidine is more complex then pyrimidine. From two pyrimidines cytosine is more complex because it has two different functional groups (amino and oxo), while uracil two the same (oxo); in the other hand, cytosine participate in the pairing process with three hydrogen bonds, and uracil with two. From two purines guanine is more complex because it contains two different functional groups (amino and oxo), while adenine only one (amino); on the other hand, guanine is involved in pairing process with three hydrogen bonds, and adenine with two. From the said it follows that from the 24 possible permutations of the four amino bases (two of Py and two of Pu type) with chemical complexity corresponds only one, the first: UCAG, the same one that one can find in the original Crick's paper (Crick, 1968, Table 1, p . 368). (In the legend of this Table Crick says that it gives the "best allocations of the 64 codons".)[14] All other permutations are related to some other important properties of GC (Damjanović, 1998; Qiu and Zhu, 2000; Yang, 2004; Damjanović and Rakočević, 2005, 2006; Dragovich & Dragovich, 2006, 2007a, 2007b).

2. For now there is no explanation of what is the meaning of this determination with the number 6. Our opinion is that it may be related to the hypothesis that perfect and friendly numbers are determinants of the genetic code (Rakočević, 1997a, pp. 60-61) (Table A.1). Among other things, the six-bit binary tree is a particular and specific, in addition to everything else, because only at it the sum of numbers in the two internal branches (two octets) corresponds to the sum of the first pair of friendly numbers (220 + 284 = 504); and each two neighboring branches give the same result (504); all together as a realization of a logical square: (0) 220 + 284 = 504; (1) 156 + 348 = 504; (2) 92 + 412 = 504; (3) 28 + 476 = 504 (Table A.2). On the other hand, the sum of the first quartet of numbers (on the binary tree "0-63") is 6, the first octet 28, the first half numbers (from 0 to 31) 496, which is a realization of the first three perfect numbers; finally, if we count all the numbers 0-63, and return back (ciclicity!) (when the number 63 becomes 64, and 0 becomes 127), we will receive as a result 8128, which is actually the fourth perfect number. Such a realization is possible only on the six-bit binary tree, not on any other. [Hint: The sum of the first four perfect numbers takes 13th place in the system of multiples of **the number 666**, which subsystem consists of multiples of the number 037, valid for the genetic code (Shcherbak, 1993, 1994, 2003, 2008; Rakočević, 2008) (Table A.1)[15].]

[14] If we calculate, for example, the positions of the golden mean, as in Table 3, but according to permutation GACU, therefore according to the latest permutation in the set of 24, then as the "golden" AAs appear: F, T, Q, N, D, E, S, R, G. Obviously, adequate chemical pairing with the "complements" is no longer possible. By this, it is important to mention that this AAs sequence, as well as that in Table 3 both follow from the standard (Table 1) as well as mitochondrial genetic code. (About differences between two codes see Comment 8).

3. Having in mind that within the Sukhodolec's system of the distribution of hydrogen atoms to amino acids (Sukhodolec, 1985) the "key" classification is realized through the numbers 5 and 8 (as here), may be presented the hypothesis that – that comes from the specific position of these numbers in the Fibonacci series (5:8 = 0.625 and 8:5 = 1.6 versus for example 8:13 = 0.61538461 ...). The Sukhodolec's system: The **5** hydrogen atoms in G; 7 in A, S, D, C; 9 in P, T, Q, H; 11 in V, F, M, Y and 13 in L, I. The **8** hydrogen atoms in N; 10 in Q, 12 in W and 14 in K,R.

4. In a first attempt (Rakočević & Jokić, 1996) we have named the „contact amino acids" as "etalonic" ("alanine stereochemical type as a measurement subject, and three other types as measurement *etalons* and measurement subject at the same time") and later (Rakočević, 2006) as "contact" AAs; contact, because the amino acid side chain ("body") is directly connected with amino acid functional group ("head"), while by AAs of alanine type this is not the case. At amino acids of this type the connection is mediated by an H-C-H group; in all except in threonine, where the connection is realized through the H-C-$CH_3$ group[16]. And, just this situation could deny our concept that four stereochemical types, as it was originally indicated by Popov (1989), are at the same time four types not only of configuration but of constitution (in the plane) too. Namely, at the first glance it appears to be the same by isoleucine as by threonine (the same "screen": H-C-$CH_3$)[17]. Yes, but the binding within threonine, as well as within all the remaining 15 AAs of alanine stereochemical type is realized through a primary carbon, and within isoleucine through the secondary one.

---

[15] „Here is wisdom. Let him that hath understanding count the number of the beast: for it is the number of a man; and his number is Six hundred threescore and six" (The New Testament, Revelation, 13: 18).

[16] Obviously, here is a hydrogen atom from the H-C-H group replaced by a methyl one.

[17] From this objective chemical similarity follows that only these two AAs (T & I) have not only enantiomers but still diastereomers.

5. By itself it is understandable that the existence of a such CIS (Cyclic Invariant System) assumes the existence of a corresponding PIS and NIS (Protein Interactive System, and Nucleotide Interactive System, respectively), as in the act of origin of the genetic code, and in all stages of evolution of the life on Earth.

6. In their certain setting, all three hypotheses on the origin of the genetic code, including the Sukhodolec's one, imply the existence of a prebiotic evolution of the genetic code only, and not the post biotic (see footnote 12); in other words, all these hypotheses imply a complete genetic code from the very beginning[18]. The so-called exceptions to the standard genetic code, represent only the degree of freedom. Unfortunately, in the current science the majority of research people is still of the opinion, that it dates from the early seventies, that the life started with an incomplete genetic code of the seven-eight, then ten, fifteen, eighteen, to be stopped - as "frozen" – to the "today's" twenty amino acids[19].

7. Within decimal numbering system there are the same-two-digit numbers: 11, 22, ..., 55, 66, ... 99; the first 11, the central 55 and the last 99. If one takes a rule: to be a change for 1 in the first, and then in second digit-position, that leads to the system, which middle case is this pattern in the genetic code (Table B.9).

8. The differences in mitochondrial code in relation to standard one are as follows: Isoleucine (I) AUU, AUC; Methionine (M) AUA, AUG; Tryptophan (W) UGA, UGG; „Stop" codons: UAA, UAG (as in standard one) and AGA, AGG, instead these last two codons to be coded for arginine (R) as in standard genetic code.

---

[18] The hypothesis on *a complet genetic code* the reader can find in our previous work (Rakočević, 2004, p. 231: „By this hypothesis, derived from presented facts as we understand them, we support the stand point that genetic code is one and unique, universal, valid for everything living, in fact, it is the condition for origin and evolution of life").

[19] Jukes, 1973, p. 24: „The idea that the number of amino acids in earlier codes was less than twenty is favoured by the fact that some of 'simple' amino acids, such as serine, glycine and alanine, have four of more codons, and that the codons may be arranged in 'quartets' ... in which only the first two bases confer specificity on the amino acid". As we now see from CIPS (Table 14) the key principle, valid for the genetic code is a *Block aufbau principle* which requires the presence of each amino acid, each codon, and each base in it, in an exact and correct position within the system itself.

9. From regularities, harmonic structures and proportions containning within mitochondrial genetic code (see Appendix C) it follows that standard and mitochondrial code posses a paralel and colective prebiotic evolution; colective in the sense of a unity and harmonized dynamic of PIS and NIS (Protein Interactiv System and Nucleotide Interaktive System).

**Acknowledgement**

I am grateful to Branko Dragovich for the indication of the Sukhodolec's works and helpful, stimulating discussion.

## **Appendix A:** Some additional relationships within three splittings

| x | 6x | (1:3)x | 66x | (11:3)x | 666x | (111:3)x |
|---|---|---|---|---|---|---|
| 27 | 162 | 9.000 | 1782 | 99.000 | 17982 | 999 |
| 26 | 156 | 8.666 | 1716 | 95.333 | 17316 | 962 |
| 25 | 150 | 8.333 | 1650 | 91.666 | 16650 | 925 |
| 24 | 144 | 8.000 | 1584 | 88.000 | 15984 | 888 |
| 23 | 138 | 7.666 | 1518 | 84.333 | 15318 | 851 |
| 22 | 132 | 7.333 | 1452 | 80.666 | 14652 | 814 |
| 21 | 126 | 7.000 | 1386 | 77.000 | 13986 | 777 |
| 20 | 120 | 6.666 | 1320 | 73.333 | 13320 | 740 |
| 19 | 114 | 6.333 | 1254 | 69.666 | 12654 | 703 |
| 18 | 108 | 6.000 | 1188 | 66.000 | 11988 | 666 |
| 17 | 102 | 5.666 | 1122 | 62.333 | 11322 | 629 |
| 16 | 96 | 5.333 | 1056 | 58.666 | 10656 | 592 |
| 15 | 90 | 5.000 | 990 | 55.000 | 9990 | 555 |
| 14 | 84 | 4.666 | 924 | 51.333 | 9324 | 518 |
| 13 | 78 | 4.333 | 858 | 47.666 | 8658 | 481 |
| 12 | 72 | 4.000 | 792 | 44.000 | 7992 | 444 |
| 11 | 66 | 3.666 | 726 | 40.333 | 7326 | 407 |
| 10 | 60 | 3.333 | 660 | 36.666 | 6660 | 370 |
| 09 | 54 | 3.000 | 594 | 33.000 | 5994 | 333 |
| 08 | 48 | 2.666 | 528 | 29.333 | 5328 | 296 |
| 07 | 42 | 2.333 | 462 | 25.666 | 4662 | 259 |
| 06 | 36 | 2.000 | 396 | 22.000 | 3996 | 222 |
| 05 | 30 | 1.666 | 330 | 18.333 | 3330 | 185 |
| 04 | 24 | 1.333 | 264 | 14.666 | 2664 | 148 |
| 03 | 18 | 1.000 | 198 | 11.000 | 1998 | 111 |
| 02 | 12 | 0.666 | 132 | 7.333 | 1332 | 074 |
| 01 | 06 | 0.333 | 66 | 3.666 | 666 | 037 |

**Table A.1**. The integer multiples of 6, 66, 666 and 037. This Table is a variant of Table 3 in Ref. (Rakočević, 2008) as well as of Table 2 in Ref. (Rakočević, 1997a, p. 61). The multiples of 6, 66 and 666 correspond to the multiples of 1/3, 11/3 and 111/3 respectively. As it is directly visible only in third step all multiples are integer, starting with Shcherbak's "Prime quantum 037". [In 6th step the "Prime quantum" is 37037, in 9th 37037037 an so on.] The red patterns in last column correspond to the patterns of nucleon number in Genetic code as follows: 10 x 111 of nucleons within 15 side chains as well as 15 "heads" of non-four-codon AAs, total 10 x 222. The 333 of nucleons within 8 side chains of four-codon AAs. The 592 (a half of third friendly number) of nucleons within 8 "heads" of four-codon AAs, in total: 333 + 592 = 925; and 592 – 333 = 259. Notice here a determination with Pythagorean law, because 333 = 9 x 37; 592 = 16 x 37 and 925 = 25 x 37. The number 8658 = 7770+088 in 13th row represents the sum of the first four perfect numbers (6+28+496+8128 = 8658). Within the set of 6, 66, 666, 6666, 66666 etc., only the number 666 corresponds to the "Prime quantum" 037, valid for the genetic code.

/00 - 07/08 - 15/16 - 23/24 - 31//32 - 39/40 - 47/48 - 55/56 - 63/

28 92 156 220 284 348 412 476

64 64 64 64 64 64 64

/00 - 07/00 - 15/00 - 23/00 - 31//00 - 39/00 - 47/00 - 55/00 - 63/

28 120 276 496 780 1128 1540 2016

92 156 220 284 348 412 476

**Table A.2.** The determination of the series of the numbers 0-63 (in correspondence with the six-bit binary tree) by perfect numbers (here visible: 28 and 496) and by the sum consisted of the first pair of the friendly numbers: 220+284 = 504 (after: Rakočević, 1997a, Figure 7, p. 60).

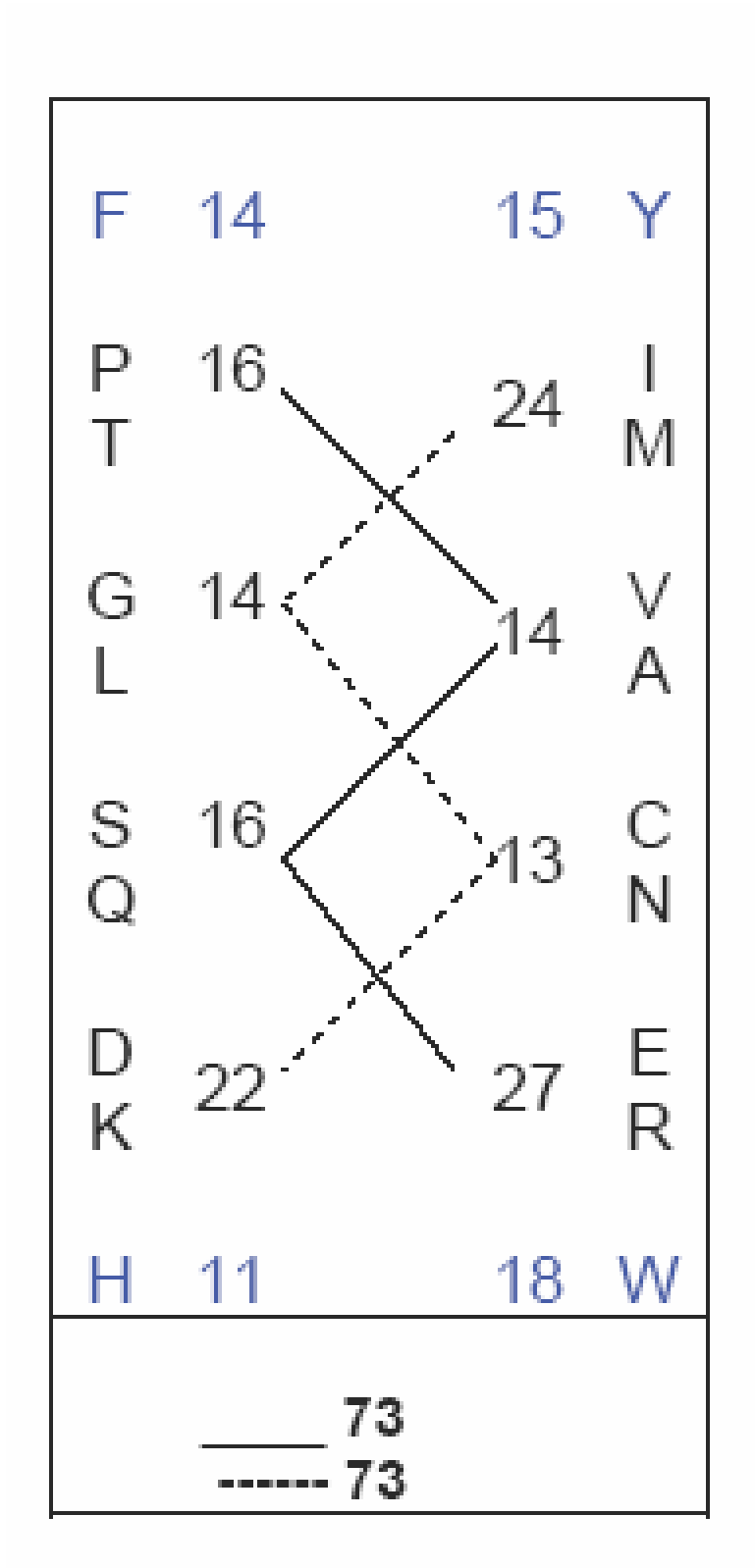

| | | | | |
|---|---|---|---|---|
| F | 14 | | 15 | Y |
| P T | 16 | (69) | 24 | I M |
| G Q | 12 | | 18 | V N |
| | | (57) | | |
| S L | 18 | | 09 | C A |
| D K | 22 | (78) | 27 | E R |
| H | 11 | | 18 | W |

**Table A.3 (left).** This Table is the same as Table 5, except three pairs of amino acids are adding, and aromatic AAs are up and down. The order follows the positions of complements on the binary-code tree (Fig. 1).

**Table A.4 (right).** The order follows the size of complement-molecules, with a new amino acid grouping. For the digital patterns of the atom number notations the principle of minimum change is valid in a specific manner: on the second position 5-6-7 and on the first one 7-8-9.

| | | | | |
|---|---|---|---|---|
| F | 14 | | 15 | Y |
| P<br>T | 16 | (29+40) | 24 | I<br>M |
| G<br>L | 14 | (28) | 14 | V<br>A |
| S<br>Q | 16 | (29) | 13 | C<br>N |
| D<br>K | 22 | (28+50) | 27 | E<br>R |
| H | 11 | | 18 | W |

**Table A.5.** This Table is the same as Table A.3 with a new amino acid grouping. The effect of the principle of minimum change as well as the principle of block-aufbau is obvious: 28-29; 40-50.

**Appendix B:** Some additional relationships within Cyclic Invariant Periodic System (CIPS) of standard genetic code

| | | | |
|---|---|---|---|
| 54 | **F** 14 | 15 **Y** | 51 |
| | **L** 13 | 04 **A** | |
| | **Q** 11 | 08 **N** | |
| | **P** 08 | 13 **I** | |
| | **T** 08 | 11 **M** | |
| 39 | **S** 05 | 05 **C** | 60 |
| | **G** 01 | 10 **V** | |
| | **D** 07 | 10 **E** | |
| | **K** 15 | 17 **R** | |
| | **H** 11 | 18 **W** | |
| Odd | **49** | **61** | |
| Even | **44** | **50** | |

**Table B.1.** Atom number within four quarters and two halves of CIPS. The effects of harmonic principles are self-evident; the multiples of number 6 also: 54 = 9 x 6 (44 = 54 – 10); 60 = 10 x 6 (50 = 60 – 10); 49 = [(8 x 6) + 1] (39 = 49 – 10); 61 = [(10 x 6) + 1] (51 = 61 – 10).

| | | | |
|---|---|---|---|
| F | 14 | 15 | Y |
| L | 13 | 04 | A |
| Q | 11 | 08 | N |
| P | 08 | 13 | I |
| T | 08 | 11 | M |
| S | 05 | 05 | C |
| G | 01 | 10 | V |
| D | 07 | 10 | E |
| K | 15 | 17 | R |
| H | 11 | 18 | W |
| F | 14 | 15 | Y |
| L | 13 | 04 | A |
| Q | 11 | 08 | N |
| P | 08 | 13 | I |
| T | 08 | 11 | M |
| S | 05 | 05 | C |
| G | 01 | 10 | V |
| D | 07 | 10 | E |
| K | 15 | 17 | R |
| H | 11 | 18 | W |

1 x 68 / 68 x 2

**Table B.2.** The number of amino acid molecules in proportion 2:3 and the number of atoms within side chains 1:2. On the right: the quantum "1 x 68" make the aliphatic "golden" AAs plus two smaller aliphatic non-complements (D & K), and the quantum "2 x 68" all others. On the left: the quantum "1 x 68" make four AAs (G, P and V, I) of non-alanine stereochemical type (larger diversity!) plus two and two AAs – two double acidic (D, E) and only two amides (N, Q), what means a very specific diversity; and the quantum "2 x 68" all other AAs.

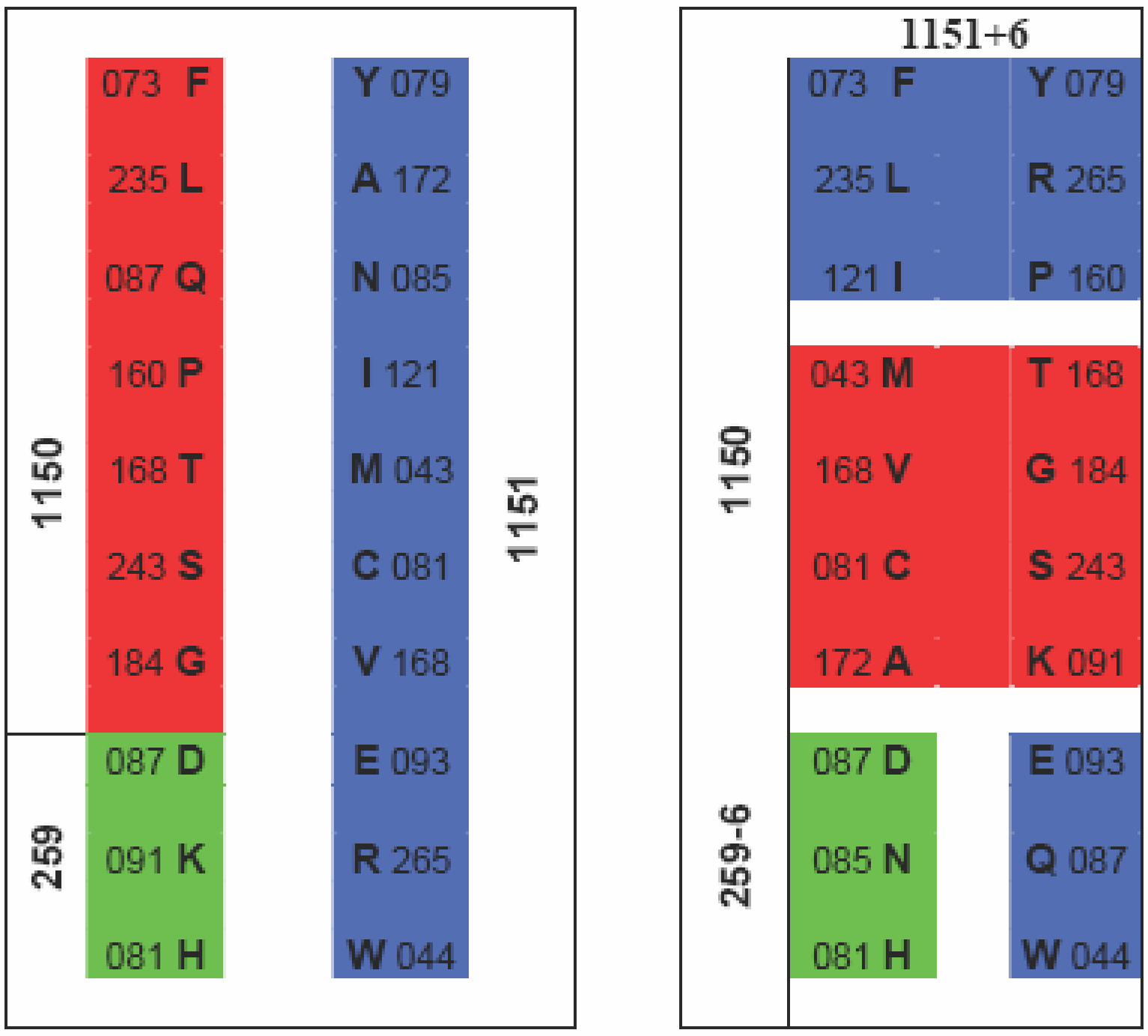

**Table B.3 (left).** The block-aufbau principle as well as the principle of self-similarity are on the scene: the quantum of 1150 atoms within coding codons (their Py-Pu bases) for seven „golden“ AAs (red block) in relation to the same such quantum valide for eight AAs in the system over there in Table B.4. At the same time the principle of minimum change is valid: 1150 versus 1151 (blue block). Notice also these relations: 1151 + 259 = 01410 („golden“ AAs versus „non-golden“ AAs in relation to 1150 : 1410). Quantum „1150“ make all seven „golden“ AAs.

**Table B.4 (right).** The self-similarity through the same patterns: the seven nonpolar AAs (F, L, I, M, V, C, A) play the role of seven “golden” AAs (F, L, Q, P, T, S, G) in previous Table (Table B.3). The polar/nonpolar AAs after (Kyte & Doolittle, 1982). The quantum „1150“ make three „golden“ AAs (T, G, S) together with three their complements (M, V, C) plus two other – A & K (Alanine as first possible hydrocarbon amino acid; lysine as a simple amine derivative – the simpler of total two: K and R).

| | | | | | |
|---|---|---|---|---|---|
| 073 | F | 91 | 107 | Y | 079 |
| 235 | L | **57** | 100 | R | 265 |
| 121 | I | 57 | 41 | P | 160 |
| 043 | M | 75 | 45 | T | 168 |
| 168 | V | 43 | 01 | G | 184 |
| 081 | C | 47 | 31 | S | 243 |
| 172 | A | 15 | 72 | K | 091 |
| 087 | D | 59 | 73 | E | 093 |
| 085 | N | 58 | 72 | Q | 087 |
| 081 | H | 81 | 130 | W | 044 |

(583 + 57)+(672+ 131) = **1443**

| 893 | 224 | 253 | 1190 |
|---|---|---|---|
| 1111+6 | | **1443** | |
| 1443 - 326 | | 1443 | |

**Table B.5.** The arrangement (as in Table B.4) into two and two areas; in the first of blue and green tones there are 1443 of atoms within coding codons (Py-Pu bases); and in second one of orange and pink tones, there are 1443 – 326 atoms. (The quantum "326" is the same as the number of atoms within nine nucleotides in three stop codons). At the same time the 23 amino acid molecules (Table 1) possess exactly 1443 of atoms. [The six-codon AAs (L,S,R) are included two times, as by Shcherbak (1994).] Notice that1443 is 1/6 of 8658 where 8658 is the sum of the first four perfect numbers (6+28+496+8128 = 8658 = 7770+0888) (cf. the 13th position within the system, presented in Table A.1.

| | | | |
|---|---|---|---|
| 073 F | Y 079 | 073 F | Y 079 |
| 235 L | A 172 | 235 L | A 172 |
| 087 Q | N 085 | 087 Q | N 085 |
| 160 P | I 121 | 160 P | I 121 |
| 168 T | M 043 | 168 T | M 043 |
| 243 S | C 081 | 243 S | C 081 |
| 184 G | V 168 | 184 G | V 168 |
| 087 D | E 168 | 087 D | E 168 |
| 091 K | R 265 | 091 K | R 265 |
| 081 H | W 044 | 081 H | W 044 |
| 1114 / 1446 | | 1115 / 1445 | |

**Table B.6** The number of molecules in proportion 1:1 on the left and 2:3 on the right. The atom number balance within coding codons (Py-Pu bases), regarding both arrangements, as 1114:1115 and 1446:1445.

| | | | | | |
|---|---|---|---|---|---|
| 87 - 1 | 14 **F** | **Y** 15 | 073 **F** | **Y** 079 | $101\underline{2}$ |
| | 13 **L** | **A** 04 | 235 **L** | **A** 172 | |
| | 11 **Q** | **N** 08 | 087 **Q** | **N** 085 | |
| | 08 **P** | **I** 13 | 160 **P** | **I** 121 | |
| 29 | 08 **T** | **M** 11 | 168 **T** | **M** 043 | 535 |
| | 05 **S** | **C** 05 | 243 **S** | **C** 081 | |
| 88+1 | 01 **G** | **V** 10 | 184 **G** | **V** 168 | $101\underline{3}$ |
| | 07 **D** | **E** 10 | 087 **D** | **E** 168 | |
| | 15 **K** | **R** 17 | 091 **K** | **R** 265 | |
| | 11 **H** | **W** 18 | 081 **H** | **W** 044 | |

**Table B.7** The atom number balance within amino acid molecules (side chains) on the left as well as within coding codons (Py-Pu bases) on the right, in relation to middle amino acid class.

| n | | a | b | AAs |
|---|---|---|---|---|
| 1 | Ribose-5-phospate | 11 | 081 | H |
| 2 | 3-Phosphoglycerate | 11 | 508 | G S C |
| 3 | Pyruvate | 27 | 575 | A L V |
| 4 | 2-Oxoglutarate | 46 | 605 | P E Q R |
| 5 | Phosphoenolpyruvate | 47 | 196 | W F Y |
| 6 | Oxaloacetate | 62 | 595 | T M I D N K |
| | Odd | 85 | [(853-1) x 1] | |
| | Even | 119 | [(853+1) x 2] | |

**Table B.8.** The six amino acid biosynthetic precursors. The precursors order (n) after number of atoms (a) within the sets of belonging AAs and the sets of belonging coding codons (b). Notice a balance in column "b" (the proportion 1:2 with a deviation for ±1) and disbalance in column "b" (85:119).

| | | | | | | |
|---|---|---|---|---|---|---|
| up : 59 → (10 x 6) − 1 | 59 | A 04 | | 15 K | 54 | up : 54 → (09 x 6) ± 0 |
| down: 43 → (7 x 6) + 1 | | V 10 | | 01 G | | down: 48 → (08 x 6) ± 0 |
| | | L 13 | | 15 R | | |
| | | F 14 | | 13 I | | |
| | (11) | W 18 | | 08 P | (11) | |
| | | H 11 | | 11 Q | | |
| | | Y 15 | | 08 N | | |
| | 48 | D 07 | | 10 E | 43 | |
| | | S 05 | | 08 T | | |
| | | C 05 | | 11 M | | |
| Odd | | 55 | (01) | 56 | (10) | 111 |
| Even | | 47 | (01) | 46 | | 93 |

**Table B.9.** As in Table B.4, with seven nonpolar AAs, the self-similarity through the same patterns is still once on the scene: the seven AAs from the odd positions in precursor system (Table B.8) (H; A, L, V; W,F,Y) play the role of seven "golden" AAs in Table B.3. The system is starting by source aliphatic AAs (A-L and K-R) with an involving of two simpler AAs from class (G-V, P-I); follow four and four AAs with a large scale of diversity; on the left four aromatic AAs, two inner (heterocyclic) and two outer ("pure" aromatic), and on the right: two more complex (N, Q) from class (D-E & N-Q) and two more complex from the class (G-V, P-I). The next, key position, keep two double acidic AAs (D,E), belonging to two classes at the same time: they belong to the class of chalcogene AAs (down) and to class with two amides. The effects of harmonic principles are self-evident, including a special algorithm (see Comment 7). Notice also a full balance for atom number (55 + 47 = 56 + 46 = 102) in contrary to disbalance (85:119) in Table B.8.

**Appendix C:** Some additional relationships within Cyclic Invariant Periodic System (CIPS) of mitochondrial genetic code

| | | | |
|---|---|---|---|
| 073 F | Y 079 | 073 F | Y 079 |
| 235 L | A 172 | 235 L | A 172 |
| 087 Q | N 085 | 087 Q | N 085 |
| 160 P | I 079 | 160 P | I 079 |
| 168 T | M 085 | 168 T | M 085 |
| 243 S | C 081 | 243 S | C 081 |
| 184 G | V 168 | 184 G | V 168 |
| 087 D | E 168 | 087 D | E 168 |
| 091 K | R 172 | 091 K | R 172 |
| 081 H | W 087 | 081 H | W 087 |
| 1409 / 1101 | | 1419 / 1091 | |

**Table C.1.** The mitochondrial genetic code in relation to the standard one (cf. Table B.6 and see Comments 8 and 9). As in standard code (Table B.6) here we have the same proportions: the number of molecules in proportion 1:1 on the left and 2:3 on the right. The atom number balance within coding codons (Py-Pu bases), regarding both arrangements, as 14**0**9:14**1**9 and 1101:1**09**1.

| | | | | | |
|---|---|---|---|---|---|
| 073 | F | 91 | 107 | Y | 079 |
| 235 | L | 57 | 15 | A | 172 |
| 087 | Q | 72 | 58 | N | 085 |
| 160 | P | 41 | 57 | I | 079 |
| 168 | T | 45 | 75 | M | 085 |
| 243 | S | 31 | 47 | C | 081 |
| 184 | G | 01 | 43 | V | 168 |
| 087 | D | 59 | 73 | E | 093 |
| 091 | K | 72 | 100 | R | 172 |
| 081 | H | 81 | 130 | W | 087 |

550 + (550+155) = 1255

1255 1255

**Table C.2.** Two columns of AAs with two and two columns of numbers; outer: number of atoms within coding codons (Py-Pu bases); inner: number of nucleons within amino acid molecules (side chains). As in Table C.1 here we have the same proportion for the number of molecules, 2:3, but the proportion for the number of atom is 1:1. But relations between atom number within codons and nucleon number within amino acid molecules is noteworthy in a special respect: the number of atom 2 x 1255 and the number of nucleons 1 x 1255 ?! (Within the columns, on the left: 550 and on the right 550+155.) If the Nature is still of a great "N", and then is – too much!

| | U | C | A | G | G | A | C | U | |
|---|---|---|---|---|---|---|---|---|---|
| F | 5 | 1 | 0 | 0 | 0 | 2 | 1 | 3 | Y |
| P | 1 | 9 | 1 | 1 | 0 | 2 | 1 | 3 | I |
| T | 1 | 5 | 5 | 1 | 1 | 3 | 0 | 2 | M |
| D | 1 | 1 | 2 | 2 | 3 | 3 | 0 | 0 | E |
| K | 0 | 0 | 5 | 1 | 5 | 1 | 5 | 1 | R |
| H | 1 | 3 | 2 | 0 | 3 | 1 | 0 | 2 | W |
| | 09 | 19 | 15 | 05 | 12 | 12 | 07 | 11 | |
| | (U+A / C+G) (24 / 24)<br>48 (8 x 6) | | | | (G+C / A+U) (19 / 23)<br>42 (7 x 6) | | | | |

**Table C.3.** The number of nucleotides within coding codons (Py-Pu bases) in blue areas of previous Table (Table C.2).

| | U | C | A | G | G | A | C | U | |
|---|---|---|---|---|---|---|---|---|---|
| **L** | 9 | 5 | 2 | 2 | 5 | 1 | 5 | 1 | **A** |
| **Q** | 0 | 2 | 3 | 1 | 0 | 4 | 1 | 1 | **N** |
| **S** | 6 | 6 | 3 | 3 | 2 | 0 | 1 | 3 | **C** |
| **G** | 1 | 1 | 1 | 9 | 5 | 1 | 1 | 5 | **V** |
| | 16 | 14 | 09 | 15 | 12 | 06 | 08 | 10 | |

(U+A / C+G) (25 / 29)
54 (9 x 6)

(G+C / A+U) (20 / 16)
36 (6 x 6)

**Table C.4.** The number of nucleotides within coding codons (Py-Pu bases) in red areas of Table C.2. Regarding two Tables (C.3 & C.4) together we see a determination with multiples of number 6 in accordance with the continuity principle: (**6** x 6), (**7** x 6), (**8** x 6) and (**9** x 6).

| | | | |
|---|---|---|---|
| 073 | F | Y | 079 |
| 235 | L | A | 172 |
| 087 | Q | N | 085 |
| 160 | P | I | |
| 168 | T | M | 085 |
| 243 | S | C | 081 |
| 184 | G | V | 168 |
| 087 | D | E | 093 |
| 091 | K | R | 172 |
| 081 | H | W | 087 |

1150 / 594 x 2 = 1188
1188+172 = 20 x 68

| | | | |
|---|---|---|---|
| 073 | F | Y | 079 |
| 235 | L | A | 172 |
| 087 | Q | N | 085 |
| 160 | P | I | |
| 168 | T | M | 085 |
| 243 | S | C | 081 |
| 184 | G | V | 168 |
| 087 | D | E | 093 |
| 091 | K | R | 172 |
| 081 | H | W | 087 |

1101 +259 = 20 x 68 = 1360
1150 +1360 = 02 x 1255

**Table C.5 (left).** The quantum of "1150" atoms within coding codons (Py-Pu bases) we had two times in standard genetic code, in Tables B.3 and B.4. The quantum "594" corresponds to the number of atoms within 61 amino acid molecules (side chains) in standard genetic code (Table 1).

**Table C.6 (right).** This Table, in relation to previous one, shows that two basic AAs (K & R) together with two heterocyclic AAs (H & W) make "a tab on the scale" in the balance.

**Appendix D.** Some harmonic structures from previous works

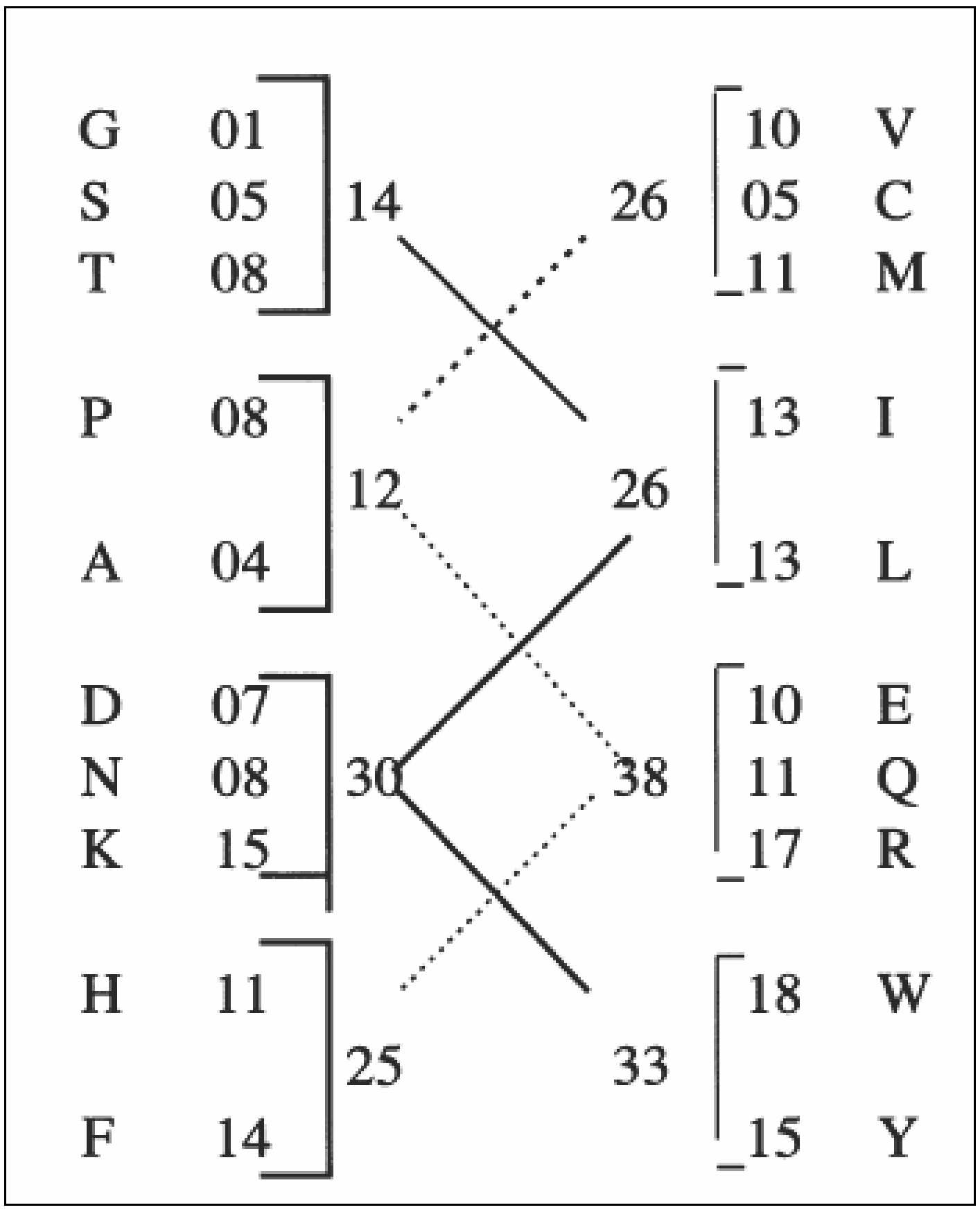

**Table D.1.** Two amino acid classes generated through the influence of two catalysts (Scheme 5 in Rakočević, 1998). On the left the smaller molecules of AAs, handled by class II of aminoacyl-tRNA synthetases, whereas on the right the larger molecules of AAs, handled by class I of synthetases. On the full line there are 102+1 and on the dotted one 102-1 of atoms. The distinct arrangement of five classes from CIPS (Table 14) is self-evident: the contact AAs (G-V and P-I) surround the chalcogene AAs (S-C and T-M). In the other hand four source aliphatic AAs (A-L and K-R) surround the class of two double acidic AAs (D & E) and their amides (N & Q).

| | | | | | | | |
|---|---|---|---|---|---|---|---|
| 28 | 09<br>19 | G P<br>A K | | 23<br>30 | V I<br>L R | 53 | **81** |
| 53 | 13<br>15<br>25 | S T<br>D N<br>F H | | 16<br>21<br>33 | C M<br>E Q<br>Y W | 70 | **123** |
| **81** | | | | | | **123** | 204 |

**Table D.2.** The arrangement in accordance to the principle: "a little" and "full" in relation to "small" and "large". So, on the left there are AAs (81 atoms) from the left side of previous Table (class II), on the right from right side (class I of AAs with 123 atoms within side chains). At the same time very up there are AAs (81 atoms) just aliphatic and nonpolar (A,V, L, I) and "a little" polar (G, P, K, R) (hydrogen and nitrogen are less polar then oxygen!); in the other hand, except aromatic and sulfur AAs, down are AAs (the row with 123 atoms) also aliphatic, but "full" polar.

| | | | | | a | b | c | d | M |
|---|---|---|---|---|---|---|---|---|---|
| D | N | A | L | → | **189** | **189** | **221** | **221+3** | 485.49 ≈ **485** |
| R | F | P | I | → | *289* | *289* | *341* | *341+0* | *585.70 ≈ 586* |
| K | Y | T | M | → | 299 | 299 | 351 | 351+2 | 595.71 ≈ 596 |
| H | W | S | C | → | *289* | *289* | *331* | *331+1* | *585.64 ≈ 586* |
| E | Q | G | V | → | **189** | **189** | **221** | **221+3** | 485.50 ≈ **485** |
| | | | | | 1255 | 1255 | 1465 | 1465+9 | 2738.04 |

**Table D.3.** Rakočević, 2004, Table 1, p. 223: "Four choices after four types of isotopes: (a) the number of nucleons within 20 AAs side chains, calculated from the first, the lightest nuclide (H-1, C-12, N-14, O-16, S-32). (b) The number of nucleons within 20 AAs side chains, calculated from the nuclide with the most abundance in the nature [the same patterns as in (a): H-1, C-12, N-14, O-16, S-32; at heavier nuclides of other bioelements the data by (a) and (b) are not the same]. (c) The number of nucleons within 20 AAs side chains, calculated from the nuclide with the less abundance in the nature (H-2, C-13, N-15, O-17, S-36); (d) The number of nucleons within 20 AAs side chains, calculated from the last, the heaviest nuclide (H-2, C-13, N-15, O-18, S-36). (M) The AAs molecule mass. Notice that (d) is greater from (c) for exactly one modular cycle (in module 9) and that total molecules mass is equal to 2 x $37^2$. Notice also that molecule mass within five rows is realized through the same logic-patterns of notations as the first nuclide, i.e. isotope."

| | | | | | |
|---|---|---|---|---|---|
| D | **E** | **Y** | S | **T** | (6 x 6)±0 |
| | | | | | (6 x 6)+1 |
| N | **Q** | *G* | C | **M** | |
| A | **L** | **F** | V | **I** | |
| K | **R** | *P* | H | **W** | |
| (66)±0 | | | (66)-1 | | |

**Table D.4.** Rakočević, 2004, Table 9, p. 229: "This system follows from the system in Table 4. First row (down): N-ended AAs. Second row: solely C-ended AAs. Last row (up): O-ended AAs. First to last row: remaining five AAs (one solely H-ended, two S-ended and two N-, O-ended, all five as a ''combination''. Within the cross there are only the exceptions: horizontally five the mentioned combining AAs; vertically: Y as aromatic within aliphatic AAs; G without carbon; F as aromatic within aliphatic AAs; and, finally, P as cyclic aliphatic amino acid. In the system there is a balanced proportionality as follows: within horizontal leg of the cross there are (6 x 6) ±0 of atoms, and within vertical leg (without glycine), there are (6 x 6) +1. Without cross: on the left there are (66) ±0 and (66)-1 on the right. [cf. this ''combination-cross'' sub-system with ''four-codon-AAs-cross'' sub-system in Codon path cube (Swanson, 1984; Rakočević, 1997b)]." Now four CIPS amino acid classes are self-evident. Notice that aromatic AAs make a "crossing" with the contact ones, but in Table B.9 only with two contact AAs (P, I) and still two amides (N, Q).

| IV | III | II | I |
|---|---|---|---|
| G-V | T-M | P-I | K-R |
| S-C | D-N | H-W | |
| Q-E | Y-F | | |
| A-L | | | |

**Table D.5.** Rakočević, 2000, Table 4, p. 281 and Rakočević, 2006a, Table 4, p. 6: "Four Amino acids classes. If amino acids pairs must be as in (Rakočević, 1998), then four classes follow from the system presented in Table 3. Non-bold: amino acids of the alanine stereochemical type, i.e. the non-etalon amino acids; bold: etalon amino acids of glycine stereochemical type (G), proline stereochemical type (P) and valine stereochemical type (V, I). About etalon and non-etalon amino acids see in: Rakočević & Jokić (1996)." Here we designated four CIPS amino acid classes in four colors.

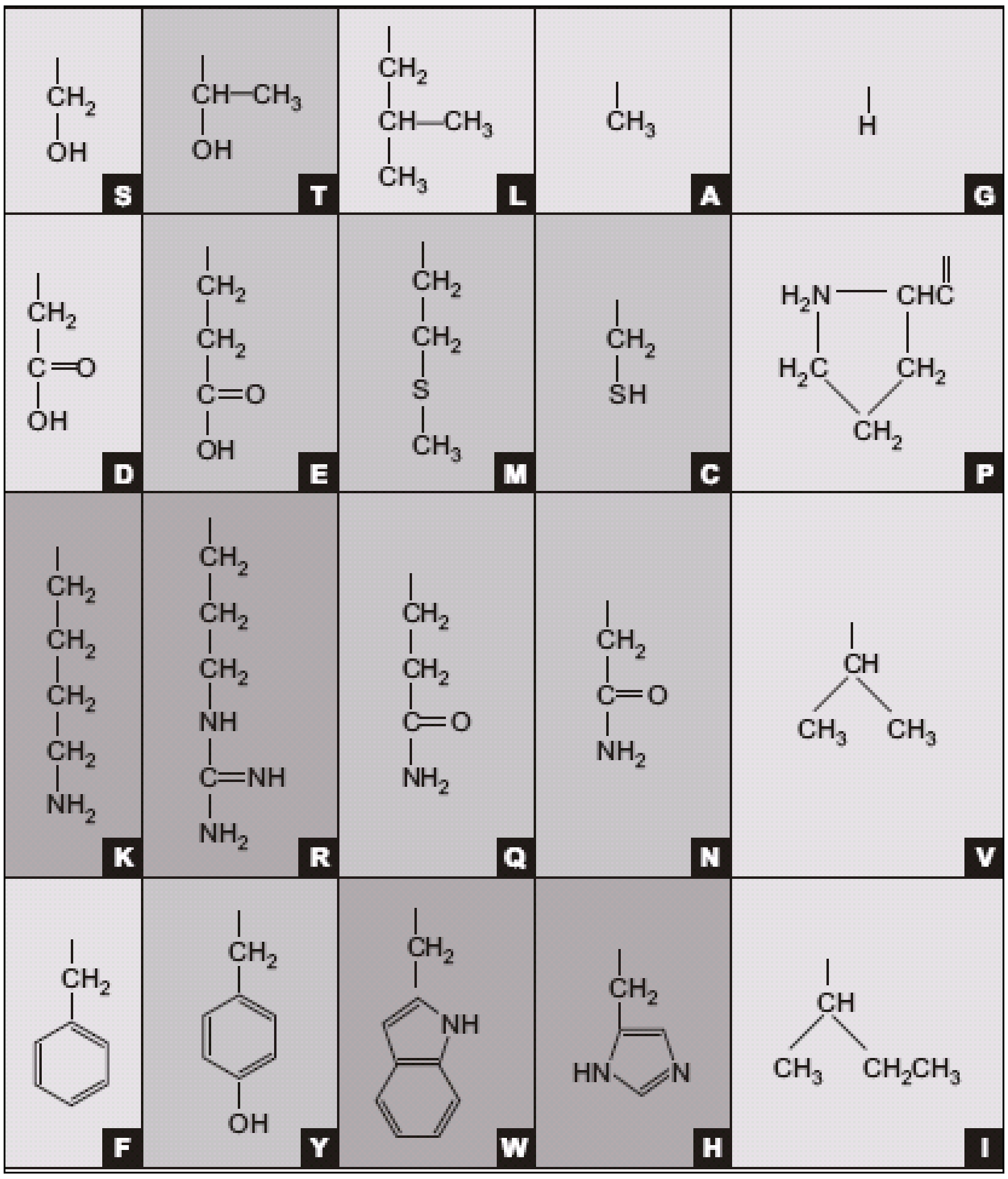

**Table D.6.1.** The arrangement AAs in accordance to Gauss' algorithm (Rakočević, 2006, Table 1.2, p. 6): "The structure of amino acid molecules. The simplest amino acid (G) … It is followed by alanine (A) whose side chain is only one $CH_3$ group … There are total of 16 amino acids of alaninic stereochemical type with one $CH_2$ group each between the "body" and the "head". The glycinic type contains glycine (G) only; valinic type contains valine and isoleucine (V, I); The last stereochemical type is prolinic with proline (P) which represents the inversion of valine in the sense that the "triangle" of three CH2 groups for the "head" is not bound by the basis, therefore not only with one but with two CH2 groups (Popov, 1989; Rakočević & Jokić, 1996). Light tones (G, P, V, I & A, L, S, D, F): invariant AAs; most dark tones (K, R, W, H): most variant AAs; less dark tones (T, E, Q, M, N, C): less variant AAs."

| | | | | |
|---|---|---|---|---|
| S | T | L | A | G |
| D | E | M | C | P |
| K | R | Q | N | V |
| F | Y | W | H | I |

**Table D.6.2.** This Table is the same one as previous, but without formulas and with a new classification in accordance to the CIPS: in orange color the chalcogene AAs (S, T & C, M), blue – double acidic and their amide derivatives (D, E & N, Q); as pink the two original aliphatic AAs with two amine derivatives (A, L & K, R), green are the "contact" AAs (G, P & V, I); and, finely, four aromatic AAs (F,Y & H, W).